\documentclass[12pt]{article}
\usepackage{amsmath}
\usepackage{epsfig}
\usepackage{latexsym}
\usepackage{a4wide}
\usepackage{bbm}
\newcommand{\I}{\text{i}}
\newcommand{\E}{\text{e}}

\newcommand{\Tr}{\text{Tr}}
\newcommand{\re}[1]{(\ref{#1})}
\renewcommand{\cite}[1]{${}^{#1}$}
\begin{document}
\abovedisplayskip17pt plus2pt minus4pt
\abovedisplayshortskip14pt plus2pt minus4pt
\belowdisplayskip17pt plus2pt minus4pt
\belowdisplayshortskip14pt plus2pt minus4pt
\title{Vacuum Polarization Using Quantum Mechanical Path Integrals}
\author{Walter Dittrich\\
  Institut f\"ur theoretische Physik, Universit\"at T\"ubingen,\\
  72076 T\"ubingen, Germany}
\maketitle
\begin{abstract}
  We compute the quantum vacuum polarization for a pure neutral scalar
  field theory within the context of single-particle quantum
  mechanics. The loop diagram is computed without ever encountering
  loop-momentum integrals. Our approach is based on standard Feynman
  path integrals. Contact is made to scalar QED.
\end{abstract}

\section{Introduction}
This paper was initiated by several recent articles on so-called
string-inspired calculations in scalar and spinor quantum
electrodynamics (QED). World-line techniques and standard quantum
mechanical path integrals were used to calculate effective actions and
loop processes without employing any loop-momentum integrals. Our
intention here is to demonstrate in a didactic and rather explicit
manner how these methods can be put to work in a simplified model
which avoids charge, spin or polarization degrees of freedom, but
brings out all the necessary physics that one would encounter in more
realistic models.

String-inspired methods in quantum field theory were first used in the
works of Bern and Kosower \cite{1}. These authors and Strassler
\cite{2} then realized that some of the well-known vacuum processes
in QED and QCD can be computed  rather easily with the aid of
one-dimensional path integrals for relativistic point
particles. Similar techniques and results can also be found in the
monograph by Polyakov \cite{3}. String-inspired methods,
particularly in QED, were then extensively studied in a series of
papers by Schmidt, Schubert and Reuter; cf., e.g., Ref. 4, where
the state of the art is reviewed extensively. There are also
contributions by McKeon \cite{5} and various co-authors who have
proved that world-line methods are extremely useful.

\section{Vacuum Polarization in a Model Field Theory\\ ${\cal
    L}'=\frac{g}{2}\, \psi^2\, \phi$}
\begin{figure}[h]
\begin{picture}(145,60)
\put(50,0){\epsfig{figure=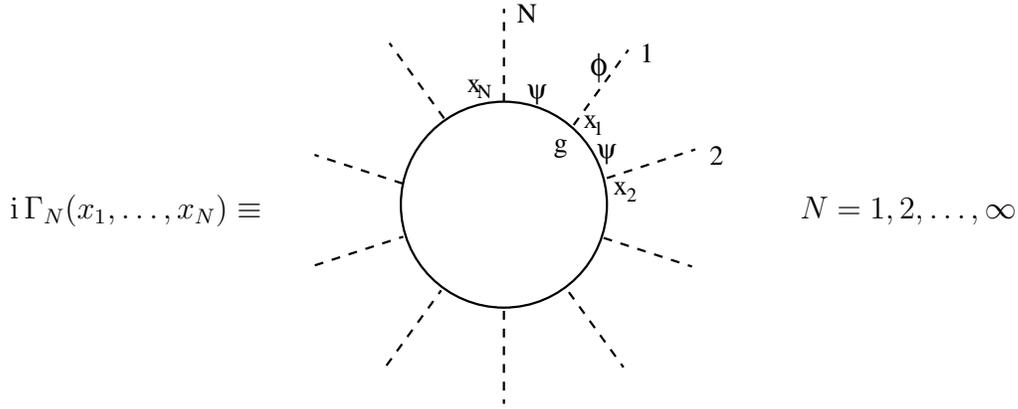,width=5.5cm}}
\put(10,25){$ \I\, \Gamma_N(x_1,\dots,x_N)\equiv$}
\put(115,25){$N=1,2,\dots,\infty$}
\end{picture}
\caption{One-loop vacuum graph in a $\psi^2\phi$-theory.}
\label{fig1}
\end{figure}

To have a comparatively simple model let us consider an interaction
Lagrangian 
\begin{equation}
{\cal L}'=\frac{g}{2} \psi^2(x)\, \phi(x), \label{1}
\end{equation}
where $g$ is the coupling constant. In the following we will be mainly
interested in the one-loop vacuum graph (Fig. \ref{fig1}).

In QED the particle circulating in the loop would be the electron
which is tied to an arbitrary number of off-shell photons. As is well
known (see, e.g., Ref. 6), loop graphs belong to a subclass of Feynman
diagrams called one-particle-irreducible diagrams. Their associated
one-particle-irreducible amplitudes $\Gamma_N(x_1,x_2,\dots,x_N)$ can
be obtained with the aid of a generating functional:
\begin{equation}
\Gamma[\phi] =\sum_{N=0}^\infty \frac{1}{N !} \int d^4x_1\,
d^4x_2\dots d^4x_N\, \Gamma_N(x_1,\dots,x_N) \, \phi(x_1)\dots
\phi(x_N). \label{2}
\end{equation}
Our interest lies with $N=2$. But for the time being we will let
$N=1,2,\dots,\infty$. Now recall from potential theory that a particle
of mass $m$ travelling to all orders in an external field $\phi(x)$ is
given by

\newpage
\begin{figure}[h]
\begin{picture}(145,30)
\put(0,0){\epsfig{figure=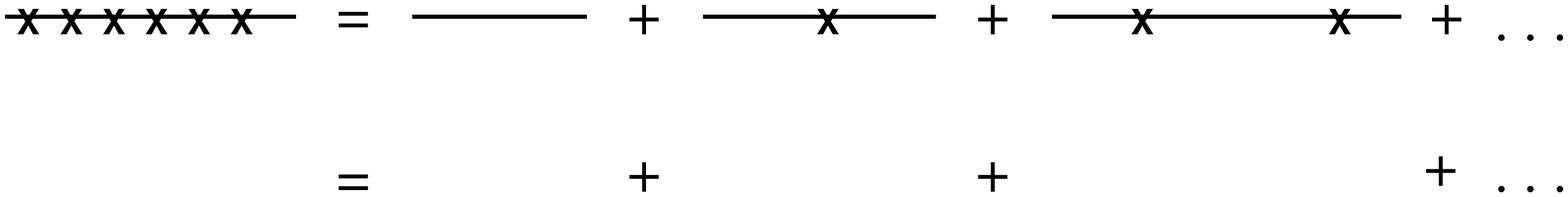,width=14.5cm}}
\put(8,1){$\Delta_+[\phi]$}
\put(43,1){$\Delta_+$}
\put(65,1){$\Delta_+\, g\phi\,\Delta_+$}
\put(97,1){$\Delta_+\, g\phi\,\Delta_+\,g\phi\, \Delta_+$}
\put(150,15){,}
\end{picture}
\end{figure}

\vspace{-3cm}

\begin{equation}
\label{3}
\end{equation}

\vspace{+1.2cm}

\noindent
where $\Delta_+$ is the Green's function of the freely propagating
$\psi$-particle ($\Delta_+\equiv \Delta_+[\phi=0]$) which satisfies
the Green's function equation $(-\partial^2+m^2) \Delta_+(x-y)
=\delta^4(x-y)$, or in momentum space, $(p^2+m^2) \Delta_+(p)=1$. Our
metric signature is $(-,+,+,+)$. Summing up the terms in the geometric
(Born-)series \re{3} we obtain
\begin{eqnarray}
\Delta_+[\phi] &=&\Delta_+\, \bigl(1+g\phi\Delta_+ + g\phi\Delta_+
g\phi\Delta_+ +\dots \bigr) \nonumber\\
&=& \Delta_+ \bigl( 1-g\phi\Delta_+ \bigr)^{-1}. \label{4}
\end{eqnarray}
It is rather trivial to rewrite Eq. \re{4} in the form
\begin{equation}
(p^2+m^2-g\phi)\, \Delta_+[\phi] =1, \label{5}
\end{equation}
or in $x$-representation 
\begin{equation}
\bigl(-\partial^2+m^2-g\phi(x)\bigr)\,
\Delta_+(x,y|\phi)=\delta^4(x-y). \label{6}
\end{equation}
Given these simple facts we can give an analytical expression for our
graph Fig. \ref{fig1}, namely,
\begin{equation}
\I \Gamma_N(x_1,\dots,x_N)=(N-1)! \frac{1}{2} \bigl(
g\Delta_+(x_1,x_2)\bigr) \bigl( g\Delta_+(x_2,x_3)\bigr) \dots \bigl(
g\Delta_+(x_N,x_1)\bigr). \label{7}
\end{equation}
The individual factors have the following origin:
\begin{itemize}
\item[(1)] The factor $(N-1)!$ takes into account that, after fixing
  one of the $\phi$-lines, a total of $(N-1)!$ topological
  inequivalent graphs can be created by permutation of the remaining
  $\phi$-lines. 

\item[(2)] One factor $g$ is assigned to each vertex.

\item[(3)] A free propagator $\Delta_+$ is assigned to each
  $\psi$-line.

\item[(4)] The factor $\frac{1}{2}$ is indicative of a neutral scalar
  field theory.
\end{itemize}
$\Gamma_N$ does not, by definition, contain external
propagators. Substituting Eq. \re{7} into \re{2} we obtain
$\Gamma[\phi]$ in one-loop approximation
\begin{eqnarray}
\I \Gamma[\phi] &=& \frac{1}{2} \sum_{N=1}^\infty \frac{(N-1)!}{N!}
\int d^4x_1\dots d^4x_N\, (g)^N\, \Delta_+(x_1,x_2)
\dots\Delta_+(x_N,x_1)\, \phi(x_1)\dots \phi(x_N) \nonumber\\
&=& \frac{1}{2} \sum_{N=1}^\infty \frac{1}{N} \int d^4x_1\dots
d^4x_N\, \bigl( g\phi(x_1)\Delta_+(x_1,x_2) \bigr) \dots 
 \bigl( g\phi(x_N)\Delta_+(x_N,x_1) \bigr)  \nonumber\\
&=&\frac{1}{2} \int d^4x_1  \sum_{N=1}^\infty \frac{1}{N} \int
d^4x_2\dots d^4x_N\, \langle x_1|g\phi\Delta_+ |x_2 \rangle \dots
\langle x_N|g\phi\Delta_+ |x_1 \rangle \nonumber\\
&=& \frac{1}{2} \int d^4x_1  \sum_{N=1}^\infty \frac{1}{N} 
 \langle x_1|(g\phi\Delta_+)^N |x_1 \rangle =\frac{1}{2} \int d^4x\,
 \langle x| \sum_{N=1}^\infty \frac{1}{N} (g\phi \Delta_+)^N |x\rangle
 \nonumber\\
&=& -\frac{1}{2} \int d^4x \langle x| \ln (1-g\phi\Delta_+) |x\rangle
=-\frac{1}{2} \Tr_x\, \ln (1-g\phi\Delta_+). \nonumber
\end{eqnarray}
Here we used the series $\ln (1-x) =-\sum_{n=1}^\infty \frac{x^n}{n}$,
$x\in (-1,+1)$. We also made use of the completeness relation $\int
d^4y\, |y\rangle\langle y| =1$ and wrote $\langle
x|g\phi\Delta_+|y\rangle =g\phi(x) \langle x| \Delta_+ |y\rangle
=g\phi(x) \Delta_+(x,y)$. So our final result reads
\begin{equation}
\I \Gamma[\phi] =\frac{1}{2}\, \Tr\, \ln (1-g\phi\Delta_+)^{-1},
\label{8}
\end{equation}
or with the aid of \re{4}:
\begin{equation}
\I\, \Gamma[\phi] =\frac{1}{2} \Tr \ln \left[
  \frac{\Delta_+[\phi]}{\Delta_+[0]} \right] \label{9}
\end{equation}
and since
\begin{displaymath}
\Delta_+[0] \equiv \Delta_+ =\frac{1}{p^2+m^2-\I\epsilon} \quad
\text{and} \qquad \Delta_+[\phi] =\frac{1}{p^2+m^2-g\phi-\I\epsilon}, 
\end{displaymath}
we have
\begin{equation}
\I\, \Gamma[\phi] =-\frac{1}{2} \Tr\, \ln
\frac{p^2+m^2-g\phi-\I\epsilon}{p^2+m^2-\I \epsilon}. \label{10}
\end{equation}
Here we employ the formula
\begin{equation}
\ln \frac{a}{b} =\int\limits_0^\infty \frac{ds}{s} \, \E^{-\I
  s(b-\I\epsilon)} -\int\limits_0^\infty \frac{ds}{s} \, \E^{-\I
  s(a-\I\epsilon)} \label{11}
\end{equation}
and obtain
\begin{equation}
\I\, \Gamma[\phi] =\frac{1}{2} \int\limits_0^\infty \frac{ds}{s} \,
\Tr\, \E^{-\I s (p^2-g\phi +m^2 -\I \epsilon)} +
\frac{1}{2} \int\limits_0^\infty \frac{ds}{s} \,
\Tr\, \E^{-\I s (p^2 +m^2 -\I \epsilon)}. \label{12}
\end{equation}
Since the last term is $\phi$-independent it is usually dropped.

Now we turn to the computation of
\begin{eqnarray}
\I\, \Gamma[\phi]&=&\frac{1}{2} \int\limits_0^\infty \frac{ds}{s} \,
\Tr_x\, \E^{-\I s (p^2-g\phi +m^2)}, \qquad m^2\equiv m^2-\I \epsilon,
  \nonumber\\
&=&\frac{1}{2} \int\limits_0^\infty \frac{ds}{s} \, \E^{-\I sm^2}\,
\int d^4 x \langle x| \E^{-\I (p^2-g\phi) s} | x\rangle. \label{13}
\end{eqnarray}
Introducing the ``Hamiltonian'' $H=p^2-g\phi$, $p_\mu=\frac{1}{\I}
\partial_\mu$, we need to calculate the trace of the quantum
mechanical transition amplitude
\begin{equation}
\langle x,s| y,0\rangle =\langle x| \E^{-\I H s} |y\rangle. \label{14}
\end{equation}
Instead of working with the Hamiltonian we now switch over to the
Lagrangian description of our system so that we can make use of
Feynman's path integral representation of Eq. \re{14}. 

Since $L=p \dot{x} -H$, $\dot{x} =\frac{\partial H}{ \partial p} =2p$:
$p=\frac{\dot{x}}{2}$, we find 
\begin{equation}
L=\frac{\dot{x}^2}{2} -\left( \frac{\dot{x}^2}{4} -g\phi \right)
=\frac{\dot{x}^2}{4} +g\phi. \label{15}
\end{equation}
Now, Feynman's path integral representation of the transition
amplitude is given by
\begin{equation}
\langle x,s| y\rangle ={\cal N} \, \int\limits_{x(0)=y,x(s)=x} {\cal
  D} x(\tau) \, \E^{\I S_{\text{cl}}[x(\tau)]}, \label{16}
\end{equation}
where 
\begin{equation}
S_{\text{cl}}[x(\tau)] =\int\limits_0^s d\tau\, L\bigl(x(\tau),
\dot{x}(\tau)\bigr), \qquad L=\frac{\dot{x}^2}{4} +g\phi. \label{17}
\end{equation}
This is all we need from Feynman's book \cite{7} or any other
monograph on Feynman path integrals in single-particle quantum
mechanics\cite{8}. Already at this stage we want to emphasize that
nowhere in the sequel do we have to compute a loop-momentum integral
as is usually required in other field theoretic approaches. 

The normalization factor ${\cal N}$ in \re{16} is determined from the
free theory, $g=0$, i.e., $H_0=p^2$. In this case we have
\begin{equation}
\langle x| \E^{-\I sp^2}|y\rangle ={\cal N}\,
\int\limits_{x(0)=y}^{x(s)=x} {\cal D}x\, \E^{\I \int\limits_0^s
  d\tau\, \frac{\dot{x}^2}{4}}. \label{18}
\end{equation}
Taking the trace on the left-hand side gives us $\bigl((dx)\equiv
d^4x\bigr)$
\begin{eqnarray} 
&&\int (dx)\, \langle x| \E^{-\I sp^2}|x\rangle =\int
  (dx)\, \langle x| \E^{-\I sp^2} \cdot \mathbbm{1} |x\rangle, \qquad
  \mathbbm{1}=\int (dp)\, |p\rangle\langle p| \nonumber\\
&=&\int (dx) \int (dp)\, \E^{-\I sp^2}\, \langle x|p\rangle \langle
  p|x\rangle, \qquad \langle x|p\rangle =\frac{\E^{\I px}}{(2\pi)^2},
  \quad \int(dx)=V_4 \nonumber\\
&=& V_4\, \int \frac{(dp)}{(2\pi)^4} \, \E^{-\I sp^2} =V_4 \left(
  \frac{-\I}{(4\pi)^2} \right) \frac{1}{s^2}. \label{19}
\end{eqnarray}
Here the four-dimensional integral over the (3+1)-dimensional momentum
space is computed as
\begin{eqnarray}
\int \frac{(dp)}{(2\pi)^4} \, \E^{-\I sp^2} &=&
\left(\int\limits_{-\infty}^\infty \frac{dp_1}{2\pi} \, \E^{-\I
    sp_1^2} \right)^3 \left(\int\limits_{-\infty}^\infty
  \frac{dp_0}{2\pi} \, \E^{-\I  sp_0^2} \right)
=\left( \frac{1}{2\pi} \left( \frac{\pi}{\I s} \right)^{1/2} \right)^3
\frac{1}{2\pi} \left( \frac{\pi \I}{s} \right)^{1/2} \nonumber\\
&=& \frac{1}{(4\pi)^2} \frac{1}{\I s^2}. \nonumber
\end{eqnarray}
The trace on the right-hand side of Eq. \re{18} is written as
\begin{displaymath}
{\cal N}\int (dx) \int\limits_{x(0)=x(s)} {\cal D}x\, \E^{\I
  \int\limits_0^s d\tau\, \frac{\dot{x}^2}{4}} \equiv {\cal N} 
\!\!\!\!\!\int\limits_{x(0)=x(s), \text{arbitrary}}\!\!\!\!\!
 {\cal D}x\, \E^{\I \int\limits_0^s d\tau\, \frac{\dot{x}^2}{4}},
\end{displaymath}
so that we end up with the useful relation
\begin{equation}
 {\cal N} \!\!\!\!\!\int\limits_{x(0)=x(s), \text{arbitr.}}\!\!\!\!\!
 {\cal D}x\, \E^{\I \int\limits_0^s d\tau\, \frac{\dot{x}^2}{4}}
=V \frac{1}{\I (4\pi)^2} \frac{1}{s^2}. \label{20}
\end{equation}
Here then is our path integral representation for the one-loop process
with an arbitrary number of external off-shell $\phi$-particle lines (
s. Eq. \re{13}):
\begin{equation}
\I\, \Gamma[\phi] =\frac{1}{2} \int\limits_0^\infty \frac{ds}{s}\,
\E^{-\I m^2 s} \,  {\cal N} \!\!\!\!\!\int\limits_{x(0)=x(s),
  \text{arbitr.}}\!\!\!\!\! {\cal D}x\, \E^{\I \int\limits_0^s
  d\tau\, \left[\frac{\dot{x}^2}{4} +g\phi\right]}. \label{21}
\end{equation}
Using a perturbative expansion of the right-hand side of Eq. \re{21}
we can write
\begin{eqnarray}
\Gamma[\phi]&=& -\frac{\I}{2} \int\limits_0^\infty \frac{ds}{s}\,
\E^{-\I m^2 s} \,  {\cal N} \!\!\!\!\!\int\limits_{x(0)=x(s),
  }\!\!\!\!\! {\cal D}x\, \E^{\I \int\limits_0^s
  d\tau\, \frac{\dot{x}^2}{4}} \sum_{N=1}^\infty \frac{(\I g)^N}{N!}
\left( \int\limits_0^s d\tau\, \phi(x(\tau))\right)^N \nonumber\\
&=& -\frac{\I}{2} \int\limits_0^\infty \frac{ds}{s}\,
\E^{-\I m^2 s} \,  {\cal N} \!\!\!\!\!\int\limits_{x(0)=x(s),
  }\!\!\!\!\! {\cal D}x\, \E^{\I \int\limits_0^s
  d\tau\, \frac{\dot{x}^2}{4}} \sum_{N=1}^\infty \frac{(\I g)^N}{N!}
\prod_{i=1}^N \int\limits_0^s d\tau_i\, \phi(x(\tau_i)) \nonumber\\
&=:& \sum_{N=1}^\infty \Gamma_N[\phi], \label{22}
\end{eqnarray}
with
\begin{equation}
\Gamma_N[\phi]=-\frac{\I}{2} \frac{(\I g)^N}{N!} \int\limits_0^\infty
\frac{ds}{s}\, \E^{-\I m^2 s} \,  {\cal N} \!\!\!\!\!\int\limits_{x(0)=x(s),
  }\!\!\!\!\! {\cal D}x\, \E^{\I \int\limits_0^s
  d\tau\, \frac{\dot{x}^2}{4}} \prod_{i=1}^N \int\limits_0^s
  d\tau_i\, \phi(x(\tau_i)). \label{23}
\end{equation}
At this point we specialize the $\phi$-field to a sum of plane waves,
\begin{equation}
\phi(x) =\sum_{i=1}^N \E^{\I k_i x}. \label{24}
\end{equation}
For $N=2$ the $\phi$-term on the right-hand side of Eq. \re{23} would
read
\begin{eqnarray}
&&\int\limits_0^s d\tau_1 d\tau_2\, \phi(x(\tau_1))\,\phi(x(\tau_2)) = 
\int\limits_0^s d\tau_1 d\tau_2\, \left( \E^{\I k_1 x(\tau_1)}+ \E^{\I
    k_2 x(\tau_1)} \right) \left( \E^{\I k_1 x(\tau_2)}+\E^{\I k_2
    x(\tau_2)} \right) \nonumber\\
&&\qquad = \int\limits_0^s d\tau_1 d\tau_2\, \left( \dots + \E^{\I k_1
    x(\tau_1)} \E^{\I k_2 x(\tau_2)}+ \E^{\I k_2 x(\tau_1)} \E^{\I k_1
    x(\tau_2)} +\dots\right)\nonumber\\
&&\qquad = 2!\int\limits_0^s d\tau_1 d\tau_2\,\E^{\I k_1
    x(\tau_1)} \E^{\I k_2 x(\tau_2)} +\dots  . \label{25}
\end{eqnarray}
Here we kept only mixed terms in $k_1$ and $k_2$, i.e., each
$\phi$-mode occurs only once.

Generalizing to $N$ we would find $N!$ instead of $2!$ in
Eq. \re{25}. This factor $N!$ then cancels the $N!$ that stands in the
denominator of Eq. \re{23}. So far we have
\begin{equation}
\Gamma_N[k_1,k_2,\dots,k_N] =(-\I) \frac{1}{2}(\I g)^N
\int\limits_0^\infty \frac{ds}{s}\, {\cal N} \int {\cal D}x\, \E^{\I
  \int\limits_0^s d\tau \left[ \frac{\dot{x}^2}{4} -m^2 \right]}
\left( \prod_{i=1}^N \int\limits_0^s d\tau_i\, \E^{\I k_i x(\tau_i)}
\right). \label{26}
\end{equation}
Introducing the ``current'' for $x$,
\begin{equation}
j(\tau)=\I \sum_{j=1}^N k_j\, \delta(\tau-\tau_j), \label{27}
\end{equation}
we can rewrite Eq. \re{26} in the form
\begin{eqnarray}
\Gamma_N[k_1,k_2,\dots,k_N] &=&(-\I) \frac{1}{2}(\I g)^N
\int\limits_0^\infty \frac{ds}{s}\, \E^{-\I m^2 s} 
\left( \prod_{i=1}^N \int\limits_0^s d\tau_i \right) 
{\cal N} \int {\cal D}x\, \E^{\I
  \int\limits_0^s d\tau  \frac{\dot{x}^2}{4}} \, \E^{\int\limits_0^s
  d\tau \,j(\tau)\, x(\tau)} \nonumber\\
 &=&(-\I) \frac{1}{2}(\I g)^N
\int\limits_0^\infty \frac{ds}{s}\, \E^{-\I m^2 s} 
\left( \prod_{i=1}^N \int\limits_0^s d\tau_i \right) 
{\cal N} \int {\cal D}x\, \E^{-\frac{\I}{4} \int\limits_0^s d\tau\, x
  \frac{d^2}{d\tau^2} x}\, \E^{\int\limits_0^s d\tau\,
  jx}. \nonumber\\
&&\label{28}
\end{eqnarray}
The operator $\frac{d^2}{d\tau^2}$, acting on $x(\tau)$ with
periodical boundary condition $x(s)=x(0)$, has zero-modes $x_0$: 
\begin{displaymath}
\frac{d^2}{d\tau^2} x(\tau) =\lambda_n\, x(\tau), \qquad x(s) =x(0),
\end{displaymath}
zero mode $x\equiv x_0=$const.: $\frac{d^2}{d\tau^2} x_0 =0$,
$\lambda_0=0$. 

\noindent
These zero modes will be separated from their orthogonal non-zero mode
partners $\xi(\tau)$ by writing $x(\tau)=x_0 +\xi$ with $\int_0^s
d\tau\, \xi(\tau)=0$, i.e., $\int_0^sd\tau\, x(\tau) =x_0$ and
\begin{equation}
\int {\cal D}x=\int d^4x_0\, \int{\cal D}\xi. \label{29}
\end{equation}
Since $x(\tau)$ (and therefore $\xi(\tau)$) is periodic we can write
\begin{equation}
x(\tau)=x_0+ \sum_{n\neq 0} c_n\, \E^{\frac{2\pi\I
    n}{s}\tau}. \label{30}
\end{equation}
The zero-mode contribution in the source term of Eq. \re{28} yields
\begin{displaymath}
\E^{\int\limits_0^s jx\,d\tau} =\E^{x_0\, \I \int\limits_0^s d\tau
  \sum_{j=1}^N k_j \, \delta(\tau-\tau_j)} =\E^{\I x_0 \sum_{j=1}^N
  k_j} 
\end{displaymath}
and therefore
\begin{equation}
\int d^4x_0\, \E^{\I x_0 \sum_{j=1}^N  k_j}  =(2\pi)^4\,
\delta^4(k_1+k_2+\dots+ k_N). \label{31}
\end{equation}
The result of combining all this information with Eq. \re{28} produces
the result
\begin{eqnarray}
\Gamma_N[k_1,\dots,k_N] &=& -\frac{1}{2} (\I g)^N (2\pi)^4
\delta\left(\sum_{j=1}^N k_j\right) \int\limits_0^\infty
\frac{ds}{(4\pi)^2 s^3}\, \E^{-\I m^2 s} \left(\prod_{i=1}^N
  \int\limits_0^s d\tau_i\right) \nonumber\\
&&\qquad\times\frac{\int {\cal D}\xi\, \E^{-\frac{\I}{4}
    \int\limits_0^s d\tau\, \xi \frac{d^2}{d\tau^2} \xi}\,
  \E^{\int\limits_0^s d\tau\, j\xi}}{  \int{\cal D}\xi\,
  \E^{-\frac{\I}{4} \int\limits_0^s d\tau\, 
    \xi \frac{d^2}{d\tau^2} \xi}}. \label{32}
\end{eqnarray}
Use has also been made of formula \re{20}:
\begin{displaymath}
{\cal N} \int d^4x_0 \int\limits_{\xi(s)=\xi(0)} {\cal D}\xi\,
  \E^{\frac{\I}{4} \int\limits_0^s d\tau\, [\dot{\xi} +\dot{x}_0]^2}
  =-\frac{\I}{(4\pi)^2} \frac{1}{s^2} V, 
\end{displaymath}
where $\dot{x}_0=0$ and the four-volume $\int d^4x_0=V$ is cancelled
on both sides. 

In Eq. \re{32} we meet the path integral
\begin{displaymath}
\int\limits_{\xi(0)=\xi(s)} {\cal D}\xi\, \E^{\I \int\limits_0^s
  d\tau\, \left(\frac{\dot{\xi}^2}{4} -\I j\xi\right)} =
  \int\limits_{\xi(0)=\xi(s)} {\cal D}\xi\, \E^{\I\, S[\xi]},
\end{displaymath}
where
\begin{equation}
S[\xi]=\int\limits_0^s d\tau\, L(\xi,\dot{\xi}), \qquad\text{and}\quad
L(\xi,\dot{\xi}) =\frac{\dot{\xi}^2}{4} -\I j\xi. \label{33}
\end{equation}
Using Eq. \re{33} we find from $\frac{d}{dt} \frac{\partial
  L}{\partial \dot{\xi}} -\frac{\partial L}{\partial \xi}=0$ the
  equation of motion,
\begin{equation}
\frac{1}{2} \frac{d^2}{d\tau^2} \xi(\tau) =-\I j(\tau), \label{34}
\end{equation}
which can be solved with the ansatz
\begin{displaymath}
\xi(\tau) =-\I \int\limits_0^sd\tau'\, G(\tau,\tau')\, j(\tau'), 
\end{displaymath}
where the Green's function equation is given by (s. later)
\begin{equation}
\frac{1}{2} \frac{d^2}{d\tau^2}\, G(\tau,\tau') =\delta(\tau-\tau')
-\frac{1}{s}. \label{35}
\end{equation}
Indeed we can write
\begin{eqnarray}
\frac{1}{2} \frac{d^2}{d\tau^2}\,\xi(\tau)&=& -\I\int\limits_0^sd\tau'
\left[ \delta(\tau-\tau') -\frac{1}{s} \right] j(\tau') =-\I j(\tau)
+\frac{\I}{s} \int\limits_0^s d\tau'\, j(\tau') \nonumber\\
&=& -\I j(\tau) +\frac{\I}{s} \int\limits_0^s d\tau'\, \I\sum_{j=1}^N
k_j\, \delta(\tau'-\tau_j) =-\I j(\tau) -\frac{1}{s} \sum_{j=1}^N
k_j \nonumber\\
&=& -\I\, j(\tau)\nonumber
\end{eqnarray}
following from the $\delta$-function in Eq. \re{32}: $\sum_{j=1}^N
k_j=0$.

We now must compute the path integrals occurring in Eq. \re{32} which
are of Gaussian type. So in an intermediate step we just consider a
single Gaussian integral:
\begin{eqnarray}
\int\frac{dx}{\sqrt{-\I \pi}}\, \E^{\I(mx^2 +vx)} &=& \int
\frac{dx}{\sqrt{-\I \pi}}\, \E^{\I m\left(x+\frac{v}{2m}\right)^2} \,
\E^{-\I \frac{v^2}{4m}} =\int\frac{dx}{\sqrt{-\I \pi}}\, \E^{\I mx^2} 
\E^{-\I \frac{v^2}{4m}} \nonumber\\
&=&\int \frac{ds\, \E^{-s^2}}{\sqrt{\pi}}\, \frac{\E^{-\I
    \frac{v^2}{4m}}}{\sqrt{m}} =1 \cdot \frac{\E^{-\I
    \frac{v^2}{4m}}}{\sqrt{m}}, \qquad
x=\frac{\sqrt{-\I}}{\sqrt{m}}. \nonumber
\end{eqnarray}
The result for a product of coupled Gaussian integrals is therefore
the generalization
\begin{equation}
\int\frac{dx_1}{\sqrt{-\I \pi}}\dots \int\frac{dx_n}{\sqrt{-\I \pi}}
\, \E^{\I \left( \sum_{l,m} x_l M_{lm} x_m + \sum_{l} x_l v_l \right)}
=\frac{\E^{-\I \sum_{l,m} v_l \frac{ (M^{-1})_{lm}}{4}
    v_m}}{\sqrt{\det M}}, \label{36}
\end{equation}
which can be proved by making a rotation on the $x$'s and $v$'s which
diagonalizes $M$ and so reduces to the one-variable case where now
$\sqrt{m}$ is replaced by $\sqrt{\det M}= \sqrt{\prod_{j=1}^n M_j}$,
where the $M_j$ are the eigenvalues of $M$. So when we first compute
the discrete version of the path integrals in Eq. \re{32} and then go
to the continuous limit, we evidently obtain (substituting
$M=-\frac{1}{4} \frac{d^2}{d\tau^2}$, $v=\frac{1}{\I} j$ in
Eq. \re{36})
\begin{eqnarray}
&&\frac{\int {\cal D}\xi\, \E^{\I \int\limits_0^s d\tau\,
    \xi\left(-\frac{1}{4} \frac{d^2}{d\tau^2}\right) \xi}\, \E^{\I
    \int\limits_0^s d\tau\,\frac{1}{\I} j\xi}}{  
\int{\cal D}\xi\, \E^{\I \int\limits_0^s d\tau\,
    \xi\left(-\frac{1}{4} \frac{d^2}{d\tau^2} \right)\xi}} \left(
    \hat{=} \frac{\E^{\I \sum_{i,j} v_i \left( \frac{1}{4M}\right)_{ij}
    v_j}}{\sqrt{\det M}} \, \sqrt{\det M} \right) \nonumber\\
&&\quad \hat{=} \E^{-\frac{\I}{2} \int\limits_0^sd\tau \int\limits_0^s
    d\tau' \sum_{i,j} j_i\, 2\left( \frac{d^2}{d\tau^2}
    \right)^{-1}_{ij} j_j}= \E^{-\frac{\I}{2}\int\limits_0^sd\tau
    \int\limits_0^s d\tau' \, j^\mu(\tau)\, G_{\mu\nu}(\tau,\tau')\,
    j^\nu(\tau')}, \nonumber
\end{eqnarray}
where
\begin{equation}
G_{\mu\nu}(\tau,\tau')=\eta_{\mu\nu} \, G(\tau,\tau'),
\qquad\text{and}\quad G(\tau,\tau') =\langle\tau| 2\left(
  \frac{d^2}{d\tau^2} \right)^{-1} |\tau'\rangle. \label{37}
\end{equation}
Now we can write Eq. \re{32} in the form
\begin{eqnarray}
\Gamma_N[k_1,\dots,k_N] &=& -\frac{1}{2} (\I g)^N (2\pi)^4
\delta\left(k_1+\dots+k_N\right)
\label{38}\\
&&\qquad \int\limits_0^\infty
\frac{ds}{(4\pi)^2 s^3}\, \E^{-\I m^2 s} \left(\prod_{i=1}^N
  \int\limits_0^s d\tau_i\right) \E^{-\frac{\I}{2}\int\limits_0^sd\tau
    \int\limits_0^s d\tau' \,  j^\mu(\tau)\, G_{\mu\nu}(\tau,\tau')\,
    j^\nu(\tau')}. \nonumber
\end{eqnarray}
Note the structure expressed in Eq. \re{38}, where loop-particle, mass
$m$, and off-shell $\phi$-particles are factorized in such a way that
the scalar particle circulating in the loop becomes multiplied by the
exponential term which is solely due to the $\phi$-particles tied to
the loop. With $j(\tau)$ given in Eq. \re{27} and 
\begin{eqnarray}
\int\limits_0^sd\tau \int\limits_0^s d\tau' \,  j\, G\,  j
&=&-\int\limits_0^sd\tau \int\limits_0^s d\tau' \,  \sum_{i,j=1}^N
k_i\, \delta(\tau-\tau_i)\, G(\tau,\tau')\, k_j\, \delta(\tau'-\tau_j)
\nonumber\\
&=& -\sum_{i,j=1}^N k_i\cdot k_j\, G(\tau_i,\tau_j), \nonumber
\end{eqnarray}
we finally obtain
\begin{equation}
\Gamma_N[k_1,\dots,k_N]\! =\! -\frac{1}{2} (\I g)^N (2\pi)^4
\delta\left(k_1+\dots+k_N\right) \int\limits_0^\infty\!\!
\frac{ds}{(4\pi)^2 s^3}\, \E^{-\I m^2 s} \prod_{i=1}^N
  \int\limits_0^s d\tau_i\, \E^{\frac{\I}{2}\sum_{i,j}^N k_i\cdot k_j\,
    G(\tau_i,\tau_j) }. \label{39}
\end{equation}
As is shown below, $G(\tau,\tau)=0=\dot{G}(\tau,\tau)$, so that there
are no terms with $k_i^2$ present, i.e., without the use of on-shell
conditions. 

Now we must devote a few lines to the Green's function of the
problem. First note that the spectrum and the eigenmodes of the
operator $\frac{\partial}{\partial \tau}$ are given by
\begin{displaymath}
\text{Spectrum}(\partial_\tau) =\I \frac{2\pi}{s} n, \qquad
\langle\tau|f_n\rangle =f_n(\tau) =\frac{1}{\sqrt{s}} \E^{\I \left(
    \frac{2\pi}{s} \right) n\tau}, \quad n\in \mathbbm{Z}, \qquad
f_n(\tau+s) =f_n(\tau). 
\end{displaymath}
One can check orthogonality, 
\begin{displaymath}
\langle f_n|f_m\rangle =\int\limits_0^s d\tau\, f_n^\star(\tau)
f_m(\tau) =\delta_{nm}, 
\end{displaymath}
and completeness,
\begin{displaymath}
\sum_{n=-\infty}^\infty f_n(\tau_2) f_n^\star(\tau_1) =\frac{1}{s}
\sum_{n=-\infty}^\infty \E^{\frac{2\pi\I}{s} (\tau_2-\tau_1)n}
=\sum_{m=-\infty}^\infty \delta(\tau_2-\tau_1-ms), 
\end{displaymath}
by Poisson's formula. Note that for $0\leq\tau_1,\tau_2\leq s$ we
obtain
\begin{displaymath}
\sum_{n=-\infty}^\infty f_n(\tau_2) f_n^\star(\tau_1)
=\delta(\tau_2-\tau_1) \qquad \text{in}\quad \int\limits_0^s
d\tau\dots \quad. 
\end{displaymath}
We are interested in the spectrum of $\partial_\tau^2$ which is given
by Spectrum$(\partial_\tau^2)=\left( \I \frac{2\pi}{s} n\right)^2=
-\frac{4\pi^2}{s^2} n^2$. 

Earlier we defined the Green's function
\begin{equation}
G(\tau_2,\tau_1) \equiv
2\langle\tau_2|(\partial_\tau^2)^{-1}|\tau_1\rangle, \label{40}
\end{equation}
which we write as
\begin{eqnarray}
2\langle \tau_2| \frac{1}{\partial_\tau^2} |\tau_1\rangle &=&
2\sum_{n=-\infty}^\infty \langle \tau_2| f_n\rangle \langle f_n|
\frac{1}{\partial_\tau^2} |f_m\rangle\langle f_m|\tau_1\rangle
\nonumber\\
&=& 2\sum_{n=-\infty}^\infty f_n(\tau_2) \frac{1}{-\frac{4\pi^2}{s^2}
  n^2} f_n^\star(\tau_1) =2s \sum_{n=-\infty, n\neq 0}^\infty
\frac{\E^{2\pi\I n \frac{\tau_2-\tau_1}{s}}}{(2\pi \I n)^2}
\label{41}\\
&=& |\tau_2-\tau_1| -\frac{(\tau_2-\tau_1)^2}{s} -\frac{s}{6}
=G(\tau_2,\tau_1). \label{42}
\end{eqnarray}
It can be easily seen that the constant $-\frac{s}{6}$ drops out in
scattering amplitude calculations, so that we can omit it at the
beginning. 

Now we are able to prove the Green's function equation mentioned
earlier in Eq. \re{35}:
\begin{eqnarray}
\frac{1}{2} \partial_\tau^2 G(\tau,\tau')
\!\!\!  &\stackrel{\re{40},\re{41}}{=}&\!\!\! \frac{1}{s}
\!\!\!\!\sum_{n=-\infty, n\neq 
  0}^\infty \E^{\frac{2\pi\I}{s} n(\tau-\tau')} =\frac{1}{s} \sum_n
  \E^{\frac{2\pi\I}{s} n(\tau-\tau')} -\frac{1}{s} =\sum_m
  \delta(\tau-\tau'-ms) -\frac{1}{s} \nonumber\\
&\stackrel{m=0}{=}& \delta(\tau-\tau') -\frac{1}{s} \qquad \text{for}
  \quad 0\leq\tau,\tau'\leq s. \nonumber
\end{eqnarray}
($\frac{1}{s}$ comes from the zero-mode, $n=0$, which is subtracted.)
With the above definition of the Green's function Eq. \re{42}, we can
easily prove the following relations:
\begin{eqnarray}
G(\tau,\tau')&=&G(\tau-\tau') =G(\tau',\tau), \qquad \text{symmetry}
\nonumber\\
\partial_\tau G(\tau,\tau')&\equiv& \dot{G}(\tau,\tau') =\text{sign}
(\tau-\tau') -\frac{2(\tau-\tau')}{s} \label{43}\\
\dot{G}(\tau,\tau')&=&-\dot{G}(\tau',\tau), \qquad
\dot{\text{}}=\frac{d}{d\tau}. \nonumber
\end{eqnarray}
Since we wanted to calculate the vacuum polarization diagram with two
$\phi$-lines we now specialize to $N=2$ in Eq. \re{39}:
\begin{eqnarray}
\Gamma_2[k_1,k_2] &=& \frac{1}{2} g^2 (2\pi)^4 \delta^4(k_1+k_2)
\int\limits_0^\infty\!\! \frac{ds}{(4\pi)^2 s^3} \, \E^{-\I m^2 s}
\int\limits_0^s d\tau_1 \int\limits_0^s d\tau_2 \, \E^{\frac{\I}{2}
  \bigl( k_1\cdot k_2 G(\tau_1,\tau_2) + k_2\cdot k_1 G(\tau_2,\tau_1)\bigr)}
  \nonumber\\
&=&  \frac{1}{2} g^2 (2\pi)^4 \delta^4(k_1+k_2)
\int\limits_0^\infty \frac{ds}{(4\pi)^2 s^3} \, \E^{-\I m^2 s}
\int\limits_0^s d\tau_1 \int\limits_0^s d\tau_2 \, \E^{\I
   k_1\cdot k_2 G(\tau_1,\tau_2)}, \label{44}
\end{eqnarray}
where we used the symmetry of $G$: $G(\tau_1,\tau_2)
=G(\tau_2,\tau_1)$.  Since $G(\tau_1,\tau_2)$ is periodic and is only
dependent on the difference $(\tau_1-\tau_2)$, the integration over
$\tau_2$ is trivial; after the $\tau_1$-integration there is no
dependence on $\tau_2$ left and hence the integration over $\tau_2$
gives just $s$. We can then choose $\tau_2=0$: 
\begin{equation}
\Gamma_2[k_1,k_2] = \frac{1}{2} g^2 (2\pi)^4 \delta^4(k_1+k_2)
\int\limits_0^\infty \frac{ds}{(4\pi)^2 s^3} \, \E^{-\I m^2 s}
\int\limits_0^s d\tau_1\, \E^{\I k_1\cdot k_2 G(\tau_1)}. \label{45}
\end{equation}
Now we use 
\begin{eqnarray}
G(\tau_1)&\equiv& G(\tau_1,0) =-\frac{\tau_1^2}{s} +\tau_1 \nonumber\\
v&=& \frac{2\tau_1}{s} -1, \qquad d\tau_1 =\frac{s}{2} dv, \qquad
\tau_1 =(v+1) \frac{s}{2} \nonumber\\
&&0\leq \tau_1\leq s, \qquad -1\leq v\leq 1, \qquad G=\frac{s}{4}
(1-v^2) \nonumber
\end{eqnarray}
and obtain 
\begin{eqnarray}
\Gamma_2[k_1,k_2] &=&\frac{1}{2} g^2 (2\pi)^4 \delta^4(k_1+k_2)
\int\limits_0^\infty \frac{ds}{2(4\pi)^2 s} \, \E^{-\I m^2 s}
\int\limits_{-1}^1 dv\, \E^{-\I k_1^2 \frac{s}{4} (1-v^2)} \label{46}\\
&=& -(2\pi)^4 \delta^4(k_1+k_2) \Pi(k^2), \nonumber
\end{eqnarray}
with 
\begin{equation}
\Pi(k^2) =-\frac{g^2}{2} \int\limits_0^\infty \frac{ds}{(4\pi)^2}
\frac{1}{s} \E^{-\I m^2 s} \frac{1}{2} \int\limits_{-1}^1 dv\, \E^{-\I
  k^2 \frac{s}{4} (1-v^2)} . \label{47}
\end{equation}
An integration by parts on the variable $v$ then produces
\begin{equation}
\Pi(k^2) =-\frac{g^2}{2(4\pi)^2} \int\limits_0^\infty \frac{ds}{s}
\E^{-\I m^2 s} + \frac{g^2}{ 2(4\pi)^2} \frac{k^2}{2} \int\limits_0^1
dv\, v^2 \frac{1}{ \left[m^2 + \frac{k^2}{4}
    (1-v^2)\right]}. \label{48}
\end{equation}
Now let us assume that the $\phi$-particle is massless; then we can
regularize $\Pi(k^2)$ at $k^2=0$ and construct $\Pi^{\text{R}}(k^2)
=\Pi(k^2) -\Pi(0)$, which is equivalent to writing Eq. \re{48} in the
simple form (we drop R again):
\begin{equation}
\Pi(k^2) =\frac{g^2}{ 2(4\pi)^2} \frac{k^2}{2} \int\limits_0^1
dv\, v^2 \frac{1}{ \left[m^2 + \frac{k^2}{4}
    (1-v^2)\right]}. \label{49}
\end{equation}
The same procedure can be carried through in Eq. \re{47}. There we
would need
\begin{eqnarray}
&&\int\limits_0^\infty \frac{ds}{s}\int\limits_0^1 dv\, \E^{-\I m^2 s}
  \left( \E^{-\I s \frac{k^2}{4} (1-v^2)} -1\right) = \int\limits_0^1
  dv   \int\limits_0^\infty \frac{ds}{s} \left( \E^{-\I s \left[ m^2 +
  \frac{k^2}{4} (1-v^2) \right]} -\E^{-\I m^2 s} \right) \nonumber\\
&&\qquad \stackrel{\re{11}}{=} - \int\limits_0^1 dv \ln \frac{m^2 +
  \frac{k^2}{4} (1-v^2)}{m^2} =-\int\limits_0^1 dv \, \ln \left( 1+
  \frac{k^2}{4m^2} (1-v^2) \right) \nonumber
\end{eqnarray}
and so obtain another version of $\Pi(k^2)$, namely,
\begin{equation}
\Pi(k^2) =\frac{g^2}{ (4\pi)^2} \frac{1}{2} \int\limits_0^1 dv\, \ln
\left( 1+ \frac{k^2}{4m^2} (1-v^2)\right). \label{50}
\end{equation}
Substituting $x=\frac{1+v}{2}$ into Eq. \re{50} we finally arrive at
\begin{eqnarray}
\Pi(k^2) &=&\frac{g^2}{ 2(4\pi)^2} \frac{1}{2} \int\limits_{-1}^1 dv\,
\ln \left( 1+ \frac{k^2}{4m^2} (1-v^2)\right) \nonumber\\
&=&\frac{g^2}{ 2(4\pi)^2} \int\limits_0^1 dx\,
\ln \left( 1+ \frac{k^2}{m^2} x(1-x)\right). \label{51}
\end{eqnarray}
It is also instructive to derive the spectral representation of our
vacuum polarization diagram. To do this we start from Eq. \re{49} and
substitute $v=\left( 1-\frac{4m^2}{M^2} \right)^{1/2}$. Our end result
is then presented as
\begin{equation}
\Pi(k^2) =k^2 \int\limits_{(2m)^2}^\infty dM^2\,
\frac{\sigma(M^2)}{k^2+M^2-\I \epsilon}, \label{52}
\end{equation}
where the so-called spectral measure is given by
\begin{equation}
\sigma(M^2) =\frac{1}{2} \left( \frac{g}{4\pi} \right)^2 \frac{ \left(
    1-\frac{4m^2}{M^2} \right)^{1/2}}{M^2}. \label{53}
\end{equation}
Finally we write for the modified massless $\phi$-particle propagator
$\bar{\Delta}_+^\phi$:
\begin{equation}
k^2\left[ 1-k^2 \int\limits_{(2m)^2}^\infty dM^2\,
\frac{\sigma(M^2)}{k^2+M^2-\I \epsilon}\right] \bar{\Delta}_+^\phi(k)
=1 \label{54}
\end{equation}
or
\begin{eqnarray}
\bar{\Delta}_+^\phi(k) &=& \frac{1}{k^2-\I \epsilon} \frac{1}{ 1- k^2
  \int\limits_{(2m)^2}^\infty dM^2\, \frac{\sigma(M^2)}{k^2+M^2-\I
  \epsilon}} \label{55}\\
&  =&\frac{1}{k^2 -\I \epsilon} +\int\limits_{(2m)^2}^\infty dM^2\,
  \frac{\sigma(M^2)}{k^2+M^2-\I   \epsilon} , \label{56}
\end{eqnarray}
where we expanded the second factor of Eq. \re{55}. We have hereby
reproduced the Lehmann-K\"allen spectral representation:
\begin{equation}
\bar{\Delta}_+^\phi(k)=\frac{1}{k^2-\I \epsilon} +\frac{1}{2} \left(
    \frac{g}{4\pi}\right)^2 \int\limits_{(2m)^2}^\infty
    \frac{dM^2}{M^2} \left( 1-\frac{4m^2}{M^2}\right)^{1/2}
    \frac{1}{k^2+M^2-\I \epsilon}. \label{57}
\end{equation}
It is interesting to compare expressions Eq. \re{53} and Eq. \re{57}
with those occurring in scalar QED. There we would find
\begin{equation}
\bar{\Delta}^\gamma_{+\mu\nu}= \left( g_{\mu\nu} -\frac{k_\mu k_\nu}{k^2}
\right) \bar{\Delta}_{+}^\gamma(k^2), \qquad \sigma(M^2) =\frac{1}{3}
\left( \frac{e}{4\pi}\right)^2 \frac{\left(1
    -\frac{4m^2}{M^2}\right)^{3/2}}{M^2}, \label{58}
\end{equation}
\begin{equation}
\bar{\Delta}_{+}^\gamma(k^2)=\frac{1}{k^2-\I \epsilon} +\frac{1}{3}
\left( \frac{e}{4\pi}\right)^2 \int\limits_{(2m)^2}^\infty
    \frac{dM^2}{m^2} \left( 1-\frac{4m^2}{M^2}\right)^{3/2}
    \frac{1}{k^2+M^2-\I \epsilon}. \label{59}
\end{equation}
Incidentally, Eq. \re{51} corresponds to the scalar QED-case:
\begin{equation}
\Pi(k^2) =\left( \frac{e}{4\pi}\right)^2 \int\limits_0^1 dx\,
(2x-1)^2\, \ln \left[ 1+ \frac{k^2}{m^2} x(1-x) \right]. \label{60}
\end{equation}

\section{Conclusion}
We have presented a one-loop calculation for a simplified model field
theory which is based on standard quantum mechanical path
integrals. We found that loop-momentum integrals can be avoided and be
replaced with simple Feynman path integrals. All the results known
from ordinary field theory can thus be obtained with much less
labor. Since our calculations have great similarity with those
occurring in QED, the reader should now be able to pursue his or her
own studies in more realistic models.

\section*{Acknowledgements}
I benefited from discussions with R. Shaisultanov, who provided the
stimulus for this investigation. I also would like to thank H. Gies
for carefully reading the manuscript.

\end{document}